\documentclass[11pt]{amsart}

\usepackage{amsmath,amsxtra,amssymb,amsthm,amsfonts}
\usepackage{mathrsfs}
\usepackage[T1]{fontenc}
\usepackage[utf8]{inputenc}
\usepackage{mathtools}   
\usepackage{tikz}
\usetikzlibrary{arrows, automata}
\usepackage{color,hyperref}
\definecolor{dblue}{rgb}{0, 0, 0.72}

\numberwithin{equation}{section}
\newtheorem{lemma}{Lemma}[section]
\newtheorem{prop}[lemma]{Proposition}
\newtheorem{theorem}[lemma]{Theorem}

\newtheorem{cor}[lemma]{Corollary}

\newtheorem{prob}[lemma]{Problem}
\newtheorem{rem}[lemma]{Remark}
\newtheorem{remark}[lemma]{Remark}

\newtheorem{definition}[lemma]{Definition}
\newtheorem{corollary}[lemma]{Corollary}

\newcommand{\re}{\begin{rem}\rm}
  \newcommand{\mar}{\end{rem}}

\newcommand{\ee }{\mathrm{I}\!\!1}

\renewcommand{\for}{\begin{eqnarray*}}

\newcommand{\mel}{\end{eqnarray*}}

\newcommand{\kl}{\pl \le \pl}
\newcommand{\gl}{\pl \ge \pl}

\newcommand{\lel}{\pl = \pl}

\newcommand{\ez}{{\mathbb E}}

\newcommand{\Mz}{{\mathbb M}}

\newcommand{\ten}{\otimes}

\DeclareMathOperator{\Ric}{Ric}

\DeclareMathOperator{\fix}{fix}

\DeclareMathOperator{\tr}{tr}

\DeclareMathOperator{\CLSI}{CLSI}
\DeclareMathOperator{\MLSI}{MLSI}

\DeclareMathOperator{\CpSI}{C_pSI}

\newcommand{\pl}{\hspace{.1cm}}

\newcommand{\qd}{\end{proof}\vspace{0.5ex}}

\newcommand{\al}{\alpha}

\newcommand{\si}{\sigma}

\newcommand{\la}{\lambda}

\newcommand{\id}{\iota_{\infty,2}^n}

\newcommand{\mm}{{\mathbb M}}

\newcommand{\pf}{\begin{proof}}

\newcommand{\be}{\left|{\atop}}

\newcommand{\xspace}{\hbox{\kern-2.5pt}}
\newcommand{\xyspace}{\hbox{\kern-1.1pt}}

\newcommand{\norm}[2]{\parallel \! #1 \! \parallel_{#2}}

\allowdisplaybreaks

\definecolor{LightGray}{rgb}{0.94,0.94,0.94}
\definecolor{VeryLightBlue}{rgb}{0.9,0.9,1}
\definecolor{LightBlue}{rgb}{0.8,0.8,1}
\definecolor{DarkBlue}{rgb}{0,0,0.6}
\definecolor{LightGreen}{rgb}{0.88,1,0.88}
\definecolor{MidGreen}{rgb}{0.6,1,0.6}
\definecolor{DarkGreen}{rgb}{0,0.6,0}
\definecolor{DarkGrreen}{rgb}{0,0.8,0}

\definecolor{VeryLightYellow}{rgb}{1,1,0.9}
\definecolor{LightYellow}{rgb}{1,1,0.6}
\definecolor{MidYellow}{rgb}{1,1,0.5}
\definecolor{DarkYellow}{rgb}{0.8,1,0.3}
\definecolor{VeryLightRed}{rgb}{1,0.9,0.9}
\definecolor{LightRed}{rgb}{1,0.8,0.8}
\definecolor{DarkRed}{rgb}{0.8,0.2,0}
\definecolor{DarkRedb}{rgb}{0.6,0.2,0}
\definecolor{DarkLila}{rgb}{0.8,0,1}
\definecolor{Beige}{rgb}{0.96,0.96,0.86}
\definecolor{Gold}{rgb}{1.,0.84,0.}
\definecolor{Goldb}{rgb}{0.7,0.3,0.5}
\definecolor{MyYellow}{rgb}{1.,0.84,0.8}

\def\11{\mathbb{I}}

\usepackage{graphicx}

\DeclareRobustCommand\openone{\leavevmode\hbox{\small1\normalsize\kern-.33em1}}

\renewcommand{\id}{\rm{id}}
\renewcommand{\be}{\begin{equation}}
	\renewcommand{\ee}{\end{equation}}
\newcommand{\bea}{\begin{eqnarray}}
	\newcommand{\eea}{\end{eqnarray}}
\newcommand{\beas}{\begin{eqnarray*}}
	\newcommand{\eeas}{\end{eqnarray*}}

\usepackage{amsmath}
\usepackage{ams math}
\usepackage{amssymb}
\usepackage{amsthm}
\usepackage{dsfont}
\usepackage{upgreek}
\usepackage{enumitem}
\usepackage{array}
\usepackage{ams fonts}
\usepackage{graphics}
\usepackage[utf8]{inputenc}
\newtheorem*{theorem*}{Theorem}
\newtheorem*{remark*}{Remark}
\newtheorem*{lemma*}{Lemma}
\newtheorem*{cor*}{Corollary}
\newtheorem*{note*}{Note}
\newtheorem*{prop*}{Proposition}
\newtheorem*{example*}{Example}

\renewcommand{\Ric}{\mbox{Ric}}

\newcommand{\Hess}{\mbox{Hess}}

\usepackage{tikz-cd}

\usepackage[margin=1.0 in]{geometry}
\begin{document}
\thanks{HL and  MJ are partially supported by NSF grants  DMS 1800872 and Raise-TAG 1839177. }
\title[Geometric approach towards $\CLSI$]{Geometric Approach Towards Complete Logarithmic Sobolev Inequalities}
\author[L. Gao]{Li Gao}
\address{Zentrum Mathematik, Technische Universit\"at M\"unchen, 85748 Garching, Germany}\email[Li Gao]{li.gao@tum.de}
\author[M. Junge]{Marius Junge}
\address{Department of Mathematics\\
University of Illinois, Urbana, IL 61801, USA} \email[Marius Junge]{mjunge@illinois.edu}

\author[H. Li]{Haojian  Li}
\address{Department of Mathematics\\
University of Illinois, Urbana, IL 61801, USA} \email[Haojian Li]{hli102@illinois.edu}

\setlength\parindent{+4ex}

\maketitle

\begin{abstract}
In this paper, we use the Carnot-Caratheodory distance from  sub-Riemanian geometry to prove entropy decay estimates for all finite dimensional symmetric quantum Markov semigroups.  This estimate is independent of the environment size and hence stable under tensorization. Our approach relies on the transference principle, the existence of $t$-designs, and the sub-Riemannian diameter of compact Lie groups and implies estimates for the spectral gap.

\end{abstract}
\section{Introduction}

Logarithmic Sobolev inequalities is a versatile tool in analysis and probability. It was first introduced by Gross \cite{Gross75a,Gross75b}, and later found rich connections to geometry, graph theory, optimal transport as well as information theory. (See e.g. \cite{BE,OV,BGL14} and the overview \cite{Ledoux} by Ledoux and by Gross \cite{Gross14}).  The natural framework of logarithmic Sobolev inequalities is given by Markov semigroups, i.e. a semigroups of measure preserving maps on a measure space. Barky-Emery theory \cite{BE}, however, indicates the importance of geometric data in obtaining good estimates.  In recent years, logarithmic Sobolev inequalities for quantum Markov semigroups have attracted a lot of attentions: see e.g. \cite{bardet,DR,DR2,KT,CM,CM18} for the connections to other functional and geometric inequalities; \cite{DR,DB14} for application in quantum information theory; \cite{Wirth,WZ,BGJ2} for infinite dimensional examples; \cite{CapelRouze,CapelRouze2} for quantum Gibbs sampler on lattice spin systems. Quantum Markov semigroups model the Markovian evolution of open quantum systems, which inevitably interact with the surrounding
environment. The motivation of this work is to study the entropy form of log-Sobolev inequlities, so-called modified log-Sobolev inequality, for finite dimensional quantum systems and their tensorization property.

A quantum Makrov semigroup on finite dimensional quantum system is described
by a Lindblad generator. Let $\Mz_n$ be the $n\times n$ matrix algebra and $\tr$ be the standard matrix trace. We consider a (symmetric) Lindlabd generator (also called Lindbladian) on $\Mz_n$
\begin{align} L(x) \lel \sum_{j=1}^k a_j^2x+xa_j^2-2a_jxa_j \label{eq:lindbald}\end{align}
where $a_j\in \Mz_n$ are self-adjoint operator. It was proved by Gorini, Kossakowski and Sudarshan \cite{GKS76} and Lindblad \cite{Lindblad76} that $L$ generates a semigroup $T_t=e^{-tL}$ of complete positive trace preserving maps, and conversely all such generators symmetric to the trace inner product has the  form \eqref{eq:lindbald}. The fixed point algebra  $N:=\{x|T_t(x)=x\pl, \pl \forall \pl t\ge 0\}$ is the commutant  $N=\{a_j|1\le j\le k\}'$ as a subalgebra. Let $E_N:\Mz_n \to N$ be the conditional expectation onto $N$, which is the projection onto the fixed point space. We say the semigroup $T_t$ or its generator $L$ satisfies $\la$-modified logarithmic Sobolev inequalities ($\la$-MLSI) for $\la>0$ if for all positive operators $\rho$,
\begin{align}\la \tr(\rho\log \rho-\rho\log E_N(\rho))\le \tr(L\rho \log \rho)\pl.\end{align}
This inequality characterizes a strong convergence property in terms of entropy that
\begin{align} D(T_t\rho||E_N(\rho))\le e^{-\la t}D(\rho||E_N(\rho))\pl, \label{eq:expdecay}\end{align}
where $D(\rho||\si)=\tr(\rho\log\rho -\rho\log \si)$ is the quantum relative entropy. In contrast to classical Markov semigroups, it is crucial to allow for environment system due to potential quantum entanglement. This leads us to consider the amplified semigroup $T_t\ten id_{\Mz_m}$ over a noiseless (finite dimensional) auxiliary systems $\Mz_m$, which goes beyond the ergodic case. We say the semigroup $T_t$ satisfies $\la$-complete logarithmic Sobolev inequalities ($\la$-CLSI) if for all $m\ge 1$, $T_t\ten id_{\Mz_m}$ satisfies $\la$-MLSI. The CLSI was first introduced in \cite{GJLfisher} and later studied in \cite{BGJ,BGJ2,WZ}. We write $\text{CLSI}(L)$ for the optimal (largest) constant $\la$ such that $T_t=e^{-Lt}$ satisfies $\la$-CLSI.
The CLSI constant governs the convergence rate independently of the size of the environment system, and more importantly, satisfies the tensorization property $\text{CLSI}(L_1\ten \id +\id\ten L_2)=\min\{\text{CLSI}(L_1),\text{CLSI}(L_2)\}$. The tensorization property was used in \cite{CapelRouze} as a key condition to obtain
size independent MLSI for quantum lattice systems. Therefore, it is desired to know whether all finite dimensional quantum Markov semigroup $T_t=e^{-Lt}$ admits $\CLSI(L)>0$.

It turns out that the above questions is closely related to matrix valued version of logarithmic Sobolev inequalities for classical Matrix semigroup. Indeed, let $G$ be a compact Lie group and its Lie algebra $\mathfrak{g}$.
Given a generating family $X=\{X_1,\cdots, X_k\}$ of the Lie algebra $\mathfrak{g}$ via Lie bracket, $X$ gives a hypoelliptic sub-Laplacian
\[ \Delta_X \lel -\sum_{j=1}^s X_jX_j \]
Given a unitary representation $u:G\to \Mz_n$, one can transfer the sub-Laplacian $\Delta_X$ to a Lindblad generator $L_X(x)=\sum_{j}-[a_j,[a_j,x]]$ where $a_j$ are self-adjoint elements such that $\pi(\exp(tX_j))=e^{ita_j}$. Then $L_X$ generates a quantum Markov semigroup $T_t=e^{-tL_X}:M_n\to M_n$, and the conditional expectation onto the fixed point subalgebra $N$ is given by 
\[E_N(x)=\displaystyle \int_G \pi(g)^*x\pi(g)d\mu(g) \pl .\] 
Here $d\mu$ is the Haar measure on $G$. $L_X$ is called a transferred Lindbladian of $\Delta_X$ via the representation $u$. Conversely, it was observed in \cite{GJLfisher} that every finite dimensional self-adjoint Lindbladian can be realized as a transferred Lindbladian from a connected compact Lie group. (We refer to Section 3 for more information on the transference principle.)

Thanks to the above transference principle, it suffices to study sub-Laplacians on compact Lie groups for matrix-valued
functions. Nevertheless, many classical tools assuming egordicity do not apply in this setting. One the technical difficulty is the fact that the generator $L_X$ is govern by a sub-Laplacian operator $\Delta_{\mathcal{X}}$. The impressive body of work by Baudoin, Thalmaier, and Grong \cite{Baudoin,GT19, BGKT19} indicates that a naive curvature identity
 \begin{align}\label{curv}
 \nabla_H \Delta_{X} \lel \hat{\Delta}_X\nabla_H  + R(\nabla_H)
 \end{align}
for some first order tensor $R$ and generator $\hat{\Delta}_X$ may fail. In fact this does not appear to hold  for the basic example $G=SU(2)$ and $L=-X^2-Y^2$ given by two out of three directions. This means that entropy decay estimates from quantum information theory have to go beyond the standard Bakry-\'{E}mry theory and circumvent the use of the famous Rothaus lemma, both are standard tools in the ergodic case. We refer to \cite{Temme,OZ} for the Rothaus lemma in the ergodic quantum case which no longer applies with additional environment. 

The main theorem of this work is a lower bound of the CLSI constant of a so-called transferred quantum Markov semigroup $T_t=e^{-L_Xt}$ via the sub-Riemannian structure of $X=\{X_1,\cdots, X_s\}$ on $G$.
\begin{theorem}\label{main} Let $G$ be a connected compact Lie group and $\mathfrak{g}$ be its Lie algebra. Let $X=\{X_1,\cdots, X_s\}$ be a family of left invariant vector field generating $\mathfrak{g}$. Suppose $\pi:G\to \Mz_n$ is unitary representation such that
\begin{equation}\label{design}
  E_N(x):= \sum_{j=1}^m \al_j
  \pi(g_j)^*x\pi(g_j)
  \end{equation}
for a finite probability distribution $\sum_{j=1}^m\al_j=1,\al_j\ge 0$. Then the CLSI constant of the transfered Lindbladian $L_X(x)=-\sum_{j=1}^s[a_j,[a_j,x]]$ satisfies
  \[ \CLSI(L_X) \gl \frac{C}{smd_X(d_X+1)^2} \pl .\]
  where $C$ is an universal constant and $d_X$ is the diameter of $G$ in the Carnot-Caratheodory distance induced by $X$.
\end{theorem}
Here the Carnot-Caratheodory distance, also called sub-Riemannian distance, is defined as 
\begin{align} d_{H}(p,q) \lel \inf_{\gamma(0)=p,\gamma(1)=q}  \int_0^1 \|\gamma'(t)\|_H dt \label{eq:subre}\end{align}
where the infimum is taken over all
piecewise smooth curves whose derivatives $\gamma'(t)$ are a.e. in the horizontal direction $H={\rm span}\{X_k(\gamma(t))\}$. This distance defines the same topology and hence $G$ admits finite diameter $d_{\mathcal{X}}$ with respect to this new metric. The equation \eqref{design} is an analog of spherical design for $G=SO(n)$ and of unitary design for $G=U(n)$, which are of interest from combinatorics and quantum computing.
Thanks to Caratheodory theorem (see \cite{Wat18}), we know that the design \eqref{design} always exists with $m\le n^2+4n+2$. Therefore, Theorem \ref{main} shows that every quantum Markov semigroup transferred from a sub-Laplacian on a compact Lie group satisfies CLSI. As a corollary, we obtain a positive solution to the existence of CLSI constants in finite dimensions.

\begin{cor}\label{cor:main} Every self-adjoint Lindbladian $L$ on a finite dimensional matrix algebra satisfies $\CLSI(L)>0$.
\end{cor}

The above results can be extended to Lindbladians $L$ satisfying GNS-symmetry of states via the noncommutative change of measure in \cite{JLRR}.
Very recently this result has been independently obtained in \cite{GR} using very different techniques. These two results are complementary: while the proof presented here requires knowledge of the Carnot-Carathodory diameter of $G$ and the size of a design for the conditional expectation and implies spectral gap, the proof by Gao and Rouz\'{e} on the other hand relies on the spectral gap and the Popa-Pimnser index \cite{pipo} of the inclusion $N\subset \Mz_m$. The lower bound in Theorem \ref{main} does not depend much on the dimension of the representation $\pi$, and holds uniformly for sub-representations of a given tensor product representations $\pi_0^{\ten k}\ten \bar{\pi}_0^{\ten_k}$. In contrast, the Popa-Pimnser index for a direct sum of irreducible representations can become  very large.

The rest of paper is organized as follows. Section 2 discusses the complete logarithmic Sobolev constant on the weighted interval. In Section 3, we use the interval result to prove Theorem \ref{main}.

\section{Complete Logarithmic Sobolev Inequalities on the Interval}
In this section we discuss the complete logarithmic Sobolev inequalities (CLSI) for the weighted interval. Let $[0,1]$ be the unit interval and $\mu$ be a probability measure on $[0,1]$. We write $L_\infty([0,1],\mu)$ (resp. $C([0,1])$ and $C^\infty([0,1])$) as the space of $L_\infty$ (resp. continuous and smooth) functions. Denote $\delta=i\frac{d}{dx}$ as the derivative operator. We shall first consider $\delta$ is a closable derivation on smooth functions $f$ with periodic boundary conditions $f(0)=f(1)$. In this case, the underlying space is equivalent to unit circle $\mathbb{T}$. We write $\delta^*$ as the adjoint operator on $L_2([0,1],\mu)$ and $\Delta_{\mu}=\delta^{*}\bar{\delta}$ as the weighted Laplacian operator.
A matrix valued function $f\in C([0,1],M_n)$ is positive if for every $t\in [0,1]$, $f(t)\ge 0$ is a positive (semi-definite) matrix. We are interested in proving the following matrix-valued modified logarithmic Sobolev inequalities that for all smooth periodic positive $f\in C^\infty([0,1],M_n)$,
\begin{align}
    \la\int_0^1 \tr(f(x)\log f(x)-f(x)\log E_\mu f)d\mu(x)\le  \int_0^1 \tr((\Delta_{\mu}f)(x)\log f(x))d\mu(x)\pl. \label{eq:interval}
\end{align}
where $E_\mu f=\int_{0}^1 fd\mu$ is the weighted mean. The left hand side above is the relative entropy $D(f||E_\mu f)$ for the matrix-valued $f$ with respect to its mean $E_\mu(f)$, and the right hand side is the Fisher information $I_{\Delta_\mu}(f)$ (also called entropy production). We denote $\CLSI([0,1],\mu)$ (resp. $\MLSI([0,1],\mu)$) for the optimal (largest) constant $\la$ such that \eqref{eq:interval} is satisfied for $n\ge 1$ and periodic positive $f\in C^\infty([0,1],M_n)$ (resp. for all periodic positive scalar valued function $f\in C^\infty([0,1])$). We also denote $\CLSI((0,1),\mu)$ (resp. $\MLSI((0,1),\mu))$ as the $\CLSI$ (resp. $\MLSI$) constant for functions $f$ without periodic boundary conditions $f(0)=f(1)$.

We emphasize that it is the constant $\CLSI([0,1],\mu)$ (or $\MLSI([0,1],\mu)$) that gives the expotential decay rate of relative entropy as in \eqref{eq:expdecay}. On the other hand, the constants $\CLSI((0,1),\mu)$ and $\MLSI((0,1),\mu))$ are not associated with a semigroup because the derivation $\delta=i\frac{d}{dx}$ are not closable without periodic boundary conditions. Nevertheless, the open interval constant $\CLSI((0,1),\mu)$ apply to more general functions and are more flexible to use with semigroups. It follows from the standard symmetrization and periodization argument in \cite[Proposition 4.5.5 \& 5.7.5]{BGL14} that the $\CLSI$ constants of $\CLSI([0,1],\mu)$ and $\CLSI((0,1),\mu)$ are related by a factor $\frac{1}{4}$ ,
\begin{align}\frac{1}{4}\CLSI([0,1], \mu)\leq \CLSI((0,1), \mu)\leq \CLSI(([0,1], \mu)\pl.\label{sp}\end{align}
It is clear that $\CLSI([0,1],\mu)\le \MLSI([0,1],\mu)$ and $\CLSI((0,1),\mu)\le \MLSI((0,1),\mu)$ but the other direction estimate is still unknown. The constant $\MLSI([0,1],\mu)$ and $\MLSI((0,1),\mu)$ for scalar-valued functions are discussed in \cite[Proposition 5.7.5]{BGL14}.


\begin{prop}\label{interval1}
Let $n$ be a positive integer and $d\mu(x)=\frac{1}{n} x^{n-1}dx$ be a probability on $[0,1]$. Then $\CLSI((0,1),\mu)\geq \frac{1}{4}\CLSI([0,1],\mu)\geq (2e^{1/2})^{-1}$ for all $n\ge 1$.
\end{prop}
\begin{proof}
Denote $dx$ as the Lebesgue measure. Consider the probability measure $d\nu^{n}(x)=\frac{1}{a_{n}}x^{n-1}e^{-\frac{x^{2}}{2}}dx$ on $[0,1]$ with $a_{n}=\int_{0}^{1}x^{n-1} e^{-\frac{x^{2}}{2}}dx$. Since for periodic boundary funtions, the underlying space is circle which has zero Ricci curvature. Then the Bakry-\'{E}mery's weighted Ricci tensor is
$$\Ric(d\nu^{n})=\Hess(\frac{x^{2}}{2}-(n-1)\ln(x))=1+\frac{n-1}{x^{2}}\geq 1$$ This implies $\CLSI([0,1], \nu^{n})\geq 2$ for funtions with periodic condition $f(0)=f(1)$. By comparing the two measures 
$\displaystyle \frac{n}{a_{n}e^{\frac{1}{2}}}\leq \frac{d\nu^{n}(x)}{d\mu(x)}\leq \frac{n}{a_{n}}$ and the change of measure in \cite[Theorem 2.14]{LJL}, we have  $\CLSI([0,1], \frac{1}{n} x^{n-1}dx)\geq 2e^{-1/2}$.
\end{proof}
More generally, we have the following criterion.
\begin{corollary}
Let $d\mu(x)=f(x)dx$ be a probability measure $[0,1]$ with second differentiable density function $f$. If there exists $k>0$ and $a>0$ such that  
$$kf^{2}(x)-f''(x)f(x)-(f'(x))^{2}\geq a>0, \quad \forall x\in (0,1).$$
Then $\CLSI((0,1),\mu)\geq \frac{1}{4}\CLSI([0,1],\mu)\geq (2e^{k})^{-1}$.
\end{corollary}
\begin{proof}
Let us consider the probability measure $d\gamma=\frac{1}{a} f(x)e^{-kx^{2}}$ with $a=\int_{0}^{1} f(x)e^{-k{x^{2}}}dx$. The weighted Ricci tensor is
$$\Ric(d\gamma)=\Hess(kx^{2}-\ln(f(x)))= \frac{2kf(x)^{2}-f''(x)f(x)-(f'(x))^{2}}{f(x)^{2}}\geq a>0 .$$ The two measures $d\mu$ and $d\nu$ are comparable 
$$\frac{ce^{-k}}{a} \leq \frac{d\nu}{d\mu}\leq \frac{c}{a}.$$
By the change of measure again, we have $\CLSI([0,1],d\mu))\geq  2e^{-k}$.
\end{proof}

\noindent The next estimate, despite of giving worse constants, applies to open interval constant and just depends on the growth order. 

\begin{prop}\label{interval}  Let $d\mu(x)\lel \frac{1}{a} h(x) dx$ be a probablity measure on $[0,1]$ and $a=\int_{0}^{1} h(x)dx$. Suppose  
 \[ c_1x^{\al}\kl h(x) \kl c_2\pl \beta x^{\beta-1} \]
for some $c_1,c_2>0$ and $0\leq \al<\beta$ with $\beta\geq 1$. Then $\CLSI((0,1), d\mu )>0$.
\end{prop} 

\begin{proof} Let $\Phi(x)=\sqrt{\frac{2}{\pi}}\int_x^{\infty} e^{-t^2/2} dt$ be the error function normalized.
Let $g:[0,\infty)\to [0,1]$ be a decreasing function such that
  \[ H(g(x)) \lel \Phi(x) \quad \text{and} \quad H(y)=\int_{0}^{y}h(x)dx.\]  Thus $H'(g(x))g'(x)=\sqrt{\frac{2}{\pi}}e^{-\frac{x^{2}}{2}}$ and 
  \begin{align}
  g'(x)=\sqrt{\frac{2}{\pi}}\frac{e^{-x^{2}/2}}{h(H^{-1}(\Phi(x)))}.
\label{ecc}
 \end{align}
  Write $\mathbb{E}f=(\int f \ dx) 1$ as the expectation to the uniform measure.
Using the fact the Gaussian measure has Ricci curvature $1$ and Lemma \ref{sp}, for positive matrix-valued function $\rho(t)=f(g(t))$ we have
 \begin{align*}
 2D(\rho\|\ez(\rho))&\kl \sqrt{\frac{2}{\pi}} \int_{0}^{\infty} \tau(\rho'(t)J_{\rho(t)}^{\log} \rho'(t)) e^{-t^2/2} dt \\
 &\lel \int_0^1 \tau(f'(x)J^{\log}_{f}f'(x))
 |g'(g^{-1}(x))|^2 h(x) dx \pl .
 \end{align*}
 where $J_{\si}^{\log}(X)=\displaystyle \int_0^\infty\frac{1}{\si+r}X\frac{1}{\si+r}dr$ is the double operator integral for $\log$ function.
By the change of variable we have $D(\rho\| \ez(\rho))=D(f\|\ez(f))$, then
it suffices to find an upper bound for $g'(x)$.
Now, our assumption $h(x)\kl c_2 \beta x^{\beta-1}$ implies $H(x)\kl c_2 x^{\beta}$ 
and hence by \eqref{ecc}
 \begin{align} h(H^{-1}(y)) \gl c(H^{-1}(y))^{\al}
 \gl c_1(\frac{y}{c_2})^{\al/\beta}  \pl .\label{eccc}\end{align}
Now, we use the inequality $\Phi(x)\gl \sqrt{\frac{2}{\pi}}\frac{e^{-x^2/2}}{x(1+x^{-2})}$ and  $\Phi(t)\gl \Phi(1)$ for $t\kl 1$. Together with \eqref{eccc}, we obtain
 \begin{align*}
 g'(x)\kl
 c^{-1}c_2^{\frac{\alpha}{\beta}}\left(\frac{2}{\pi}\right)^{\frac{1}{2}(1-\frac{\alpha}{\beta})} \left(x(1+\frac{1}{x^{2}})\right)^{\frac{\alpha}{\beta}}e^{-\frac{x^{2}}{2}(1-\frac{\alpha}{\beta})}.
 \end{align*}
Note that for small $t$ we may replace $t(1+1/t^2)$ with a constant.
Thus for $0\kl \al<\beta$, this term is bounded, and hence its square is also bounded.
\end{proof}

For $h(x)=\frac{1}{n}x^{n-1}$, the $\CLSI$ constant from above is of the order $n^2$. Nevertheless, Lemma \ref{interval} help us understand measures whose density functions $h$ do not have desired smooth ($C^{2}$) properties as in Proposition \ref{interval1}. 

\begin{rem}[CLSI constant for uniform measure]{\rm The above Proposition \ref{interval1} gives a tight constant then \cite[Example 4.7]{LJL} that $$ \CLSI((0,1),dx)\geq \frac{1}{4}\CLSI([0,1], dx)\geq \frac{1}{5}\pl, $$
which was obtained by comparing the uniform measure with a another modified Gaussian distribution. A sharper constant
$$\CLSI( (0,1), dx)\geq \frac{1}{4}\CLSI([0,1], dx)\geq \frac{\pi^{2}}{\ln 3}$$
was obtained in \cite[Theorem 4.12]{BGJ} using heat kernel estimate and monotonicity of Fisher information. Note for the scalar case $\MLSI([0,1], dx)=4\pi^2$ and $\MLSI((0,1), dx)=\pi^2$.
The method in \cite{BGJ} also applies to the weighted measure but the heat kernel estimate for the weighted Laplacian $\Delta_\mu$ is less explicit.}
\end{rem}

\begin{remark}[Extension to piecewise smooth functions]\rm{Here we discuss some subtlety about the domain of $\Delta_{\mu}$ and of the modified log-Sobolev inequality. On one hand, the semigroup $T_t=e^{-t\Delta_{\mu}}$ is defined for all functions $f\in L_1((0,1),\mu)$. $\MLSI((0,1),\mu)\geq \la$ is equivalent to that for any density function $\rho\in L_1((0,1),\mu), \rho\ge 0$ and $\displaystyle \int \rho d\mu =1$,
\[D(T_t\rho\|1)\le e^{-\la t} D(\rho\|1)\pl.\]
where $\displaystyle D(f\|1)=\int f\log f d\mu$ is the entropy functional. 
For smooth $\rho$, we can take derivative at $t=0$ and obtain 
the modified log-Sobolev inequality
\[ \la D(\rho\|1)\le I_{\Delta_{\mu}}(\rho):=\int \Delta_\mu \rho \log \rho d\mu. \]
This inequality can be extended to piecewise smooth $\rho$ where the Fisher information $I_{\Delta_{\mu}}(\rho)$ has to be interpreted as Dirichelet form
\begin{align} I_{\Delta_{\mu}}(\rho)=\lim_{n\to \infty}\mathcal{E}(\rho, f_n(\rho)):=\lim_{n\to \infty}\int \frac{d}{dx}(\rho) \frac{d}{dx}(f_n(\rho)) d\mu  \label{fisher}\end{align}
where $f_n(t)=\max\{\min\{\log(t), n\},-n\}$ is the truncated logarithmic function (see \cite[Definition 5.17]{Wirth}). Suppose $\rho:[0,1]\to \mathbb{R}$ is continuous piecewise smooth and its derivative $\rho'$ is defined and continuous except for finite points in $[0,1]$. For our purpose, it suffices to consider $\rho$ is bounded from below (also bounded from above by continuity). Thus $I(\rho)=\mathcal{E}(\rho, f_m(\rho))$ for some finite $m$.
Let $\epsilon_n$ be an approximation identity of smoothing kernels. Take $\rho_n=\rho*\epsilon_n$ by convolution, and $\rho_n'=\rho'*\epsilon_n$ be the derivative of $\rho_n$. It is readily to see that
\[\norm{\rho_n-\rho}{2}\le \norm{\rho_n-\rho}{\infty}\le \norm{\rho'_n-\rho}{1}\le \norm{\rho'_n-\rho}{2}\to 0\pl, \]
which means that both $\rho_n\to \rho$ and $\rho_n'\to \rho{'}$ in $L_2$. Hence $\rho\in  \text{dom}(\frac{d}{dx})=\text{dom}(\Delta_{\mu}^{1/2})$ by closable extension and by the Leibniz rule
$f_m(\rho)$ also in $\text{dom}(\frac{d}{dx})$. Thus the Fisher information $I_{\Delta_{\mu}}(\rho)$ is also well-defined. Note that 
by data processing inequality $\rho\mapsto \rho*\epsilon_n$ and lower-semicontinuity of relative entropy,
\[ \limsup_n D(\rho_n\|1)\le D(\rho\|1)\le \liminf_n D(\rho_n\|1)\pl.\]
For the Fisher information, 
\[ I_{\Delta_{\mu}}(\rho)= \mathcal{E}(\rho, f_m(\rho))=\lim_{n}\mathcal{E}(\rho_n, f_m(\rho_n))=\lim_{n}I_{\Delta_{\mu}}(\rho_n) \pl.\]
Here we use \cite[Corollary 7.5]{DPWS} for the continuity  $\norm{\delta(f_m(\rho_n))-\delta(f_m(\rho))}{2}\to 0$.
Thus for continuous, piecewise smooth and strictly positive $\rho$,
\[ \la D(\rho||E(\rho))\le I_{\Delta_{\mu}}(\rho)\pl. \]
The same argument works for the matrix-valued functions. For smooth matrix-valued  density $\rho:[0,1]\to \Mz_n$,
\[I_{\Delta_{\mu}}(\rho)=\int_0^1 \tr\Big( (\Delta_\mu \ten \id_n\rho)(x) \log \rho(x)\Big)dx=\int_0^1 \tr\Big(\rho'(x)J^{\ln}_{\rho(x)}\rho'(x)\Big)dx\pl.\]
where $J_\si(X)=\displaystyle \int_0^\infty\frac{1}{\si+r}X\frac{1}{\si+r}dr$ is the double operator integral for $f(x,y)=\frac{\log x-\log y}{x-y}$.
Indeed, this is clear for smooth $\rho_n=\rho* (\epsilon_n 1_{M_n})$ (entry-wise mollification) and by limit
\[ I_{\Delta_{\mu}}(\rho)=\lim_{n} I_{\Delta_{\mu}}(\rho_n) =\lim_{n}\int_0^1 \tr(\rho_n'(x)J^{\ln}_{\rho_n(x)}\rho_n'(x))dx=\int_0^1 \tr(\rho'(x)J^{\ln}_{\rho(x)}\rho'(x))dx.\]
Here we use the fact $\rho_n'\to \rho'$ in $L_2([0,1],M_n)$ and $\rho_n\to \rho$ in $\norm{\cdot}{\infty}$ (because $\rho$ is continuous). To sum up, our discussion above justifies that modified log-Sobolev inequalites 
\[\la D(\rho || E_\mu \rho)\le I_{\Delta_{\mu}}(\rho)= \int_0^1 \tau(\rho'(x)J^{\ln}_{\rho(x)}\rho'(x))dx\]
extends to piecewise smooth, strictly positive matrix-valued density function.}
\end{remark}




\section{Complete Logarithmic Sobolev Inequalities On Matrix Algebras }
In this section we prove that every symmetric quantum Markov semigroup $T_t=e^{-tL}:\Mz_{n}\to \Mz_{n}$ on Matrix algebra satisfies complete logarithmic Sobolev inequality. A quantum Markov semigroup $T_t:\Mz_n\to \Mz_n$ is a continuous family of maps satisfying
\begin{enumerate}
    \item[i)] for each $t\ge 0$, $T_t$ is completely positive and unital $T_t(1)=1$.
    \item[ii)] for any $t,s\ge 0$, $T_t\circ T_s=T_{t+s}$ and $T_0=\id_{\Mz_n}$.
\end{enumerate}
where $\id_{\Mz_n}$ is the identity map on $\Mz_n$. We denote $L_2(M_n)$ as the Hilbert-Schmidt space equipped with the inner product $\langle a,b \rangle=\tr(a^{*}b)$.
We say a semigroup $T_t$ is symmetric  if for each $t\ge 0$, $T_t$ is a self-adjoint map on $L_{2}(\Mz_{n}).$ Namely, for any $x, y\in \Mz_n$,
\[\tr(T_t(x)^*y)=\tr(x^*T_t(y))\pl.\]
The generator of the semigroup (also called Lindbladian) is a operator on $L_2(\Mz_n)$ defined as
\[\pl Lx=\lim_{t\to 0} \frac{1}{t}(x-T_t(x))\pl, \pl T_t={e^{-tL}}\pl,\]
where $L$ is a operator on $L_2(\Mz_n)$. In most of our discussion, we restrict ourselves to the  symmetric cases. Thanks to \cite{GKS,Lindblad76}, the generator of symmetric semigroups is given by 
\begin{align}L(x)=-\sum_{k=1}^{s}[a_k,[a_k,x]]=\sum_{k=1}^{s}(a_{k}^{2}x+xa_{k}^{2}-2a_{k}xa_{k}). \label{eq:lindblad}\end{align}
where $a_{k}\in M_n$ are some self-adjoint operators. Then, $L$ admits a $*$-preserving derivation given by
\begin{align*}
\delta: \Mz_{n}\to  \bigoplus_{j=1}^s\Mz_{n}, \delta(x)= i[a_{1},x]\oplus i[a_{2},x] \oplus\cdots \oplus i[a_{s},x].
\end{align*}
Recall that $\delta$ is called a derivation because it satisfies the Leibniz rule $\delta(xy)=\delta(x)y+x\delta(y)$. In particular, $L=\delta^*\delta$. The fixed-point algebra is  \[N:=\{x\in \Mz_n \pl |\pl T_t(x)=x\pl, \forall t\ge 0  \}=\{x\in \Mz_n \pl |\pl Lx=0\pl \}=\{a_{1},\dots, a_{s}\}'\pl. \] We denote by $E_{N}$ be the conditional expectation onto $N_{\fix}$.

For two states $\rho$ and $\si$ with $\tr(\rho)=\tr(\si)$, the relative entropy is defined as
\begin{align*}
D(\rho\|\si)=\begin{cases}
\tr(\rho\ln\rho-\rho\ln\si), & \mbox{if } \text{supp}(\rho)\subseteq \text{supp}(\si) \\
+\infty, & \mbox{otherwise},
\end{cases}
\end{align*}
where $\text{supp}(\rho)$ (resp. $\text{supp}(\si)$) is the support projection of $\rho$ (resp. $\si$). The Fisher information (also called entropy production) is $I(\rho)=\tr(L\rho\log \rho)$.
\begin{definition}
We say $T_t$ satisfies $\la$-modified logarithmic Sobolev inequalities ($\la$-MLSI) for $\la>0$ if for all state $\rho$
\[2\la D(\rho\|E_N(\rho))\le I(\rho)\pl.\]
We say $T_t$ satisfies $\la$-complete logarithmic Sobolev inequalities ($\la$-CLSI) for $\la>0$ if for all $m\ge 1$, $T_t\ten \id_{M_m}$ satisfies $\la$-MLSI.
\end{definition}

\begin{rem} As a matter of simplicity the results in this work is stated for environments given by matrix algebras. As the proof will show the auxiliary matrix algebra $\Mz_m$ can be replaced by any finite von Neumann algebra $\mathcal{M}$ with a specified trace.   
\end{rem}

It was proved in \cite{GJLfisher} that a symmetric quantum Markov semigroup on matrix algebra is always a transference of a classical Markov semigroup on a compact Lie group  with sub-Laplacian as the generator. Recall that for a Riemannian manifold $M$,
a \emph{H\"{o}rmander system} is a finite family of vector fields  $X=\{X_1,...,X_s\}$ such that for some global constant $l_X$, the set of iterated commutators (no commutator if $k=1$)
 \[ \bigcup_{1\le k\le l_X} \{ [X_{j_1},[X_{j_2},...,[X_{j_{k-1}},X_{j_k}]]] \pl | \pl 1\le j_1,\cdots,j_k\le s\}  \]
spans the tangent space $T_p M$ at every point $p\in M$. 
\noindent We denote $\Delta_X \lel \sum_{j=1}^s X_j^*X_j$ as the sub-Laplacian where $X_j^*$ is the adjoint operator of $X_j$ with respect to $L_2(M,\mu)$ and $\mu$ is the volume form of $M$. 

\begin{lemma}[\cite{GJLfisher}]
Let $T_t:\Mz_n\to \Mz_n$ be a symmetric quantum Markov semigroup. There exists a connected compact Lie group $G$, a unitary representation $u:G\to \Mz_n$ and a H\"ormander system $X=\{X_1,\cdots, X_d\}$ of a compact connected Lie group $G$ such that the following diagram commute
\begin{equation}\label{eq:ccss}
 \begin{array}{ccc} L_\infty(G,\mu;\Mz_n)\pl\pl &\overset{S_t\ten \id_{\Mz_n}}{\longrightarrow} & L_\infty(G,\mu;\Mz_n) \\
                    \uparrow \pi   & & \uparrow \pi  \\
                     \Mz_n \pl\pl &\overset{T_t}{\longrightarrow} & \Mz_n
                     \end{array} \pl .
                     \end{equation}
where $L_\infty(G,\mu;\Mz_n)$ denotes the matrix-valued function on $G$ and
$\pi:\Mz_n\to  L_\infty(G,\mu;\Mz_n)$ denotes the transference map $\pi(x)(g)=u(g)^*xu(g).$
\end{lemma}
We briefly describe the construction, as it will be used in later discussion (See \cite[Lemma 4.10, 5.1]{GJLfisher} for detailed proof). Let $\{a_1,\cdots, a_r\}$ be the self-adjoint elements in the \eqref{eq:lindblad}. Denote $\mathfrak{u}_m=i(\Mz_n)_{s.a.}$ as the Lie algebra of the unitary group $U(\Mz_m)$. Then $X=\{ia_1,\cdots, ia_r\}$ generates a Lie subalgebra $\mathfrak{g}$ of $\mathfrak{u}_m$ which by basically Lie's second theorem (see also \cite[Lemma 4.10]{GJLfisher}) is the Lie algebra of connected compact Lie group $G$. Let $u$ be the unitary representation induced by the Lie algebra
embedding $\mathfrak{g}\subset\mathfrak{u}_m$. One can show that the $*$-homomorphism $\pi:\Mz_n\to L_\infty(G,\mu;\Mz_n)$ satisfies that 
\[X_j\ten id_{\Mz_n}(\pi(x))\lel -i \pi([a_j,x])\pl, \Delta_X\ten id_{\mm_m}(\pi(\rho)) \lel \pi(L(\rho))\pl,\]
which yields the intertwining relation of the semigroups \eqref{eq:ccss}. 

From the above intertwining relation, we can view $T_t$ as a sub-semigroup for the matrix valued semigroup
$S_t\ten \id_{\Mz_n}$. In particular, when $t\to \infty$, we have the commutation relation for the conditional expectations
\[ (\mathbb{E}_{G}\ten \id_{\Mz_n})\circ \pi=\pi\circ E_{N}\pl,\]
where $\displaystyle \mathbb{E}_{G}f=(\int_G f d\mu) 1_G$ is the expectation on $G$ and $\mu$ be the normalized Haar measure over $G$. In particular, the conditional expectation onto the fixed-point subalgebra $N=u(G)'$ 
is given by 
\[E_{N}(\rho)=\int_{G} u(g)^{*}\rho u(g)d\mu(g)\pl.\]
By Carath\'eodory's Theorem (see e.g. Proposition 4.9 in \cite{Wat18}), there exist finitely many elements $\{g_{j}\}_{j=1}^{m}\subset G$ such that for every $\rho\in \Mz_{n}$, 
\begin{align}E_N(\rho)=\sum_{j=1}^{m}\alpha_{j}u(g_{j})^{*}\rho u(g_{j}), \label{eq:cd}\end{align}
where $\{\alpha_j\}$ is a finite probability distribution s.t. $\sum \alpha_{j}=1$ and $\alpha_{j}\geq   0$.
Furthermore, we have $\CLSI(T_{t})\geq \CLSI(S_{t})$.
This transference does not apply to $\MLSI$ because $T_t$ is a restriction of the matrix-valued amplification $S_t\ten \id_n$.
\noindent We are now ready to prove the main theorem of this paper.

\begin{theorem}\label{lindblad}
Let $T_t=e^{-tL}:\Mz_n\to \Mz_n$ be symmetric quantum Markov semigroup. Suppose $L$ is a transferred Lindbladian of a sub-Laplacian $\Delta_X$ on a connected compact Lie group $G$ given by the H\"ormander system on $X=\{X_1,\cdots, X_s\}$. Then
  \[ \CLSI(L_X) \gl \frac{C}{smd_X(d_X+1)^2} \pl .\]
  where $C_X$ is some constant depending on $X$ and $d_X$ is the diameter of $G$ in the Carnot-Caratheodory distance induced by $X$.
\end{theorem}
\begin{proof}Recall $d_H$ be the horizontal distance defined in \eqref{eq:subre} and denote $d_X:=\sup_{g_1,g_2\in G}d_H(g_1,g_2)$ as the horizontal diameter. Without loss of generality, we can always assume that $X=\{X_1,...,X_s\}$ forms orthonormal set with respect to the Riemmannian metric . Namely, for any point $g\in G$ and $\la_k\in \mathbb{R}$,
 \[  |\sum_k \la_k X_k|_g^2 =\sum_k |\la_k|^2\pl.\] 
Let $\{\al_j\}$ be the probability given in \eqref{eq:cd}.
Define recursively $\beta_1=\al_1+d_{CC}(g_1,g_2)$ and for $1\le j\le m-1$, $\beta_{j+1}=\beta_j+\al_j+d_{H}(g_j,g_{j+1})$. ($m+1$ is viewed as $1$.) Then 
 \[ \beta_m\kl \sum_{j=1}^m \al_j+ \sum_j d_{H}(g_j,g_{j+1})\kl 1+m d_X \pl .\]
We split the interval \[I_j=[\beta_j,\beta_{j+1}]=[\beta_j,\beta_{j}+\al_j]\cup [\beta_{j}+\al_j,\beta_{j+1}]:=I_j(1)\cup I_j(2)\] into intervals of length $|I_j(1)|=\al_j$ and $|I_j(2)|=d_{H}(g_j,g_{j+1})$. 
Consider the new transference map $\pi: M_{n}\to \ell_{\infty}^m(M_{n})$ defined by 
\begin{align*}
\pi(\rho)(j)=u(g_{j})^{*}\rho u(g_{j})\pl.
\end{align*}
Let  $\mathbb{E}_{\mu}(f)=\sum_{j=1}^m \al_jf(j)$ be the expectation on  $\ell_{\infty}^m$.  Then we have
 \begin{align*}
 E_{N}(\rho) &= \mathbb{E}_{\mu}(\pi(\rho))\pl ,\\
 D(\rho||E_N(\rho)) &= \tr(\rho\log \rho)-\tr(\rho\log \mathbb{E}_\mu(\pi(\rho))  \lel  D(\pi(\rho)\|\mathbb{E}_\mu(\pi(\rho)) \pl .
 \end{align*}
Let $\gamma_{j}:[0,d_{H}(g_j,g_{j+1})]\to G$ be a piecewise smooth \emph{horizantal} path such that 
\begin{align*}\gamma_{j}(0)=g_{j}\pl ,\pl  \gamma_{j}(d_{H}(g_j,g_{j+1}))=g_{j+1}\pl ,\pl  \gamma_{j}'(t)\in T_{\gamma_{j}(t)}H\pl, 
\end{align*}
for a.e. $t\in(0,d_{H}(g_j,g_{j+1})$ and $|\gamma'(t)|=1$ of unit length in the Riemmannian metric.
Then we define the piecewise  smooth matrix-valued function $\tilde{\rho}:M_m\to L_\infty([0,\beta_m],M_n)$,
\begin{align*}
\tilde{\rho}(t)=\begin{cases} u(g_{j})^{*}\rho u(g_{j}), &\quad \text{if} \quad t\in I_j(1)\\ u(\gamma_{j}(t-\beta_j-\al_j))^{*}\rho u(\gamma_{j}(t-\beta_j-\al_j)) , &\quad \text{if} \quad t\in I_{j}(2).
\end{cases}
\end{align*}
Denote $\mathbb{E}\rho= (\int \rho dt) 1$ as the mean and take $\tilde{\sigma}:=\mathbb{E}\tilde{\rho}$.
By the chain rule \cite[Lemma 3.4]{JLRR} and the non-negativity of the relative entropy
\begin{align} 
D(\pi(\rho)\|\tilde{\sigma})=D(\pi(\rho)\|\mathbb{E}\pi(\rho))+D(\mathbb{E}\pi(\rho)\|\tilde{\sigma})\ge 
D(\pi(\rho)\|\mathbb{E}\pi(\rho))
\label{eq2}
\end{align}
Moreover 
\begin{align}
D(\pi(\rho)\|\tilde{\sigma})=\sum_{j=1}^{m} \alpha_{j} D(u(g_{j})^{*}\rho u(g_{j})\|\tilde{\sigma})
\leq  D(\tilde{\rho}\|\tilde{\sigma}). \label{eq3}
\end{align}
Let $C:=\CLSI([0,\beta_m],dx)$ be the $\CLSI$ constant on the interval $[0,\beta_m]$. 
We obtain that 
\begin{align*}
D(\tilde{\rho}\|\tilde{\sigma})=D(\tilde{\rho}\|\mathbb{E}\tilde{\rho})\leq \frac{1}{C}I(\tilde{\rho})=\frac{1}{C} \int_{0}^{\beta_m}\tr\left( \tilde{\rho}'(t) J^{\ln}_{\tilde{\rho}(t)}(\tilde{\rho}'(t))  \right)
dx\end{align*}
Let $X_{t}=\gamma_{j}'(t-\beta_j)\in T_{\gamma_{j}(t-\beta_j)}H$ and $g_{t}=\gamma_{j}(t-\beta_j)$.
Denote $\phi_u:\mathfrak{g}\to i(\Mz_{n})_{s.a.}$ be the Lie algebra homomorphism induced by $u$.
For $t\in I_{j}(2)$, we have 
\begin{align*}
\tilde{\rho}'(t)=&\frac{d}{ds}\left(u(g_{t+s})^{*}\rho u(g_{t+s}) \right)|_{s=0}\\
=&u(g_{t})^{*}(-\phi_u(X_t)\rho+\rho \phi_u(X_t)) u(g_{t})\\
=& u(g_{t})^{*} (-i Y_{t}\rho+i\rho Y_{t}) u(g_{t})\\
=&-u(g_{t})^{*}(i[
Y_{t},\rho])u(g_{t}).
\end{align*}
where $Y_t=i\phi_u(X_t)$ are self-adjoint operators. Then we observe that 
\begin{align*}
I(\tilde{\rho})=&
\int_{\bigcup_j I_j(2)} \tr\left( \tilde{\rho}'(t) J^{\log}_{\tilde{\rho}(t)}(\tilde{\rho}'(t))  \right) dt \pl. 
\end{align*}
and for each $j$ and $t \in I_j(2)$, 
\begin{align*}
 \tr\left( \tilde{\rho}'(t) J^{\log}_{\tilde{\rho}(t)}\tilde{\rho}'(t)\right)
=&\int_{0}^{\infty}\tr \Big( \gamma_{j}(t-\beta_j)^{*} i[Y_{t}, \rho] \gamma_{j}(t-\beta_j) (\gamma_{j}(t-\beta_j)^{*}\rho \gamma_{j}(t-\beta_j)+r)^{-1}  \\  & \pl\pl\pl\pl\pl\pl\pl\pl \cdot \gamma_{j}(t-\beta_j)^{*} i[Y_{t}, \rho]\gamma_{j}(t-\beta_j) (\gamma_{j}(t-\beta_j)^{*}\rho \gamma_{j}(t-\beta_j)+r)^{-1} \Big) dr\pl, \\
=&\int_{0}^{\infty}\tr \left( i[Y_{t}, \rho]  (\rho +r)^{-1}  i[Y_{t}, \rho](\rho +r)^{-1}\right) dr
\end{align*}
Note that $\phi_u(X_{k})=a_k$. Suppose that for each $t\in I_j(2)$, $Y(t)=\sum_{k=1}^{s}\lambda_{k}(t) a_{k}$ and hence
$X_t=\sum_{k=1}^{s} \lambda_{k}(t) X_{k}|_{\gamma_{j}(t-\beta_j)}$. The horizontal path $\gamma_j$ has constant unit speed $|X_t|_{\gamma_j(t)}\le 1$ and hence $\sum_k \la_k(t)^2\le 1$.  Take $\omega_{r}^{k}=(\rho+r)^{-\frac{1}{2}}i[a_{k},\rho](\rho+r)^{-\frac{1}{2}}$. We have
\begin{align*} 
I(\tilde{\rho})=&\sum_{j=1}^{m}\int_{I_{j}(2)}\int_{0}^{\infty}\tr \left(  i[Y_{t}, \rho]  (\rho +r)^{-1}   i[Y_{t}, \rho] (\rho +r)^{-1}\right) dr dt
\\= & \sum_{j=1}^{m}\int_{I_{j}(2)}\int_{0}^{\infty} \sum_{k,l=1}^s\la_k(t)\la_l(t)\tr \left(  i[a_k, \rho]  (\rho +r)^{-1}   i[a_l, \rho] (\rho +r)^{-1}\right) dr dt
\\= & \sum_{j=1}^{m}\int_{I_{j}(2)}\int_{0}^{\infty} \sum_{k,l=1}^s\la_k(t)\la_l(t)\tau(\omega_{r}^k\omega_{r}^l) dr dt
\\ \le & \sum_{j=1}^{m}\int_{I_{j}(2)}\int_{0}^{\infty} s(\sum_k \lambda_k(t)^2)\sum_{k=1}^s\tr(\omega_{r}^k\omega_{r}^k) dr dt
\quad \text{(by Cauchy-Schwarz inequality)}\\
=&s|\sum_j d_{CC}(g_j,g_{j+1})| \int_{0}^{\infty}\sum_{k=1}^s\tr (\omega_{r}^k\omega_{r}^k) dr\\=&
smd_X I_L(\rho)
\end{align*}
Combining the estimates above, 
we have 
\begin{align*}
D(\rho\|E(\rho))=D(\pi(\rho)\|\mathbb{E}\pi(\rho))\le D(\pi(\rho)\|\mathbb{E}\tilde{\rho})\leq D(\tilde{\rho}\|\mathbb{E}\tilde{\rho}) \le \frac{1}{C} I(\tilde{\rho})\le \frac{smd_X}{C}
 I_L(\rho)\pl.
\end{align*}
Note that the constant $C=\CLSI([0,\beta_m],dt)=\beta_m^{-2}\CLSI([0,1],dt)$. Thus we prove the CLSI constant $\CLSI(L_X)= \frac{\CLSI([0,1],dx)}{smd_X(d_X+1)^2}$.
Finally, for non-orthonormal linearly independent $X_k'$s, the change of basis adds another multiplicative constant and completes the proof.
\end{proof}

Using the noncommutative change of measure in \cite[Theorem 4.1]{JLRR}, we obtain the existence of CLSI constant for all finite dimensional quantum Markov semigroup satisfies detailed balance condition (see e.g. \cite{JLRR} for detailed defintion.).
\begin{corollary}
Let $T_t=e^{-Lt}:\Mz_n\to \Mz_n$ be a quantum Markov semigroup GNS-symmetric with respect to a full rank state $\si$. Then $T_t$ satisfies the $\la$-$\CLSI$ constant for some $\la>0$.
\end{corollary}
\begin{proof}By \cite[Theorem 4.1]{JLRR}, we know the optimal CLSI constant of every GNS-symmetric semigroup $T_t$ is comparable to the optimal CLSI constant of a trace symmetric semigroup $\tilde{T}_t$.
\end{proof}
\noindent The argument in this section applies to complete Beckner inequalities $\CpSI$, see \cite{Li20} for definitions of $\CpSI$.
\begin{corollary}
Let $T_t=e^{-Lt}:\Mz_n\to \Mz_n$ be a quantum Markov semigroup GNS-symmetric with respect to a full rank state $\si$. Then $T_t$ satisfies the $\la$-$\CpSI$ constant for some $\la>0$.
\end{corollary}

\bibliographystyle{alpha}

\bibliography{tube}

\end{document}